\documentclass[12pt,a4paper]{article}
\usepackage{amsfonts}
\newcommand{\e}{\mathbf{e}}

\newcommand{\RR}{\mathbb{R}}
\newcommand{\CC}{\mathbb{C}}
\newcommand{\ZZ}{\mathbb{Z}}

\newcommand{\bw}{{\textstyle\bigwedge}}

\newcommand{\arr}[1]{\smash{\mathop{\longrightarrow}\limits^{#1}}}
\newcommand{\one}{1\kern-3pt{\rm l}}

\begin{document}
\begin{center}
  {\bf\Large Extra Dimensions: Will their Spinors\\ 
             Play a Role in the Standard Model?}\\[10mm]
   G.\ Roepstorff\\
   Institute for Theoretical Physics\\
   RWTH Aachen\\
   D-52062 Aachen, Germany\\
   e-mail: roep@physik.rwth-aachen.de
\end{center}
\vspace{10mm}
{\bf Abstract}. As the title suggests we try to answer a question
pertinent to models of fundamental fermions in a world of high dimension.
As will be shown, ten extra dimensions are needed to accommodate quarks
and leptons of one generation in a single spinor space. We thus propose to 
think of a spinor as a direct product of a Dirac spinor with another spinor,
naturally associated with the Riemannian geometry of the 
internal space. One attractive feature is avoidance of mirror fermions.
Along the way we encounter the combined spin group $SL(2,\CC)\times
\mbox{Spin}(10)$, possibly a symmetry group for unification theories beyond 
the Standard Model. Finally, we present arguments to support a special
choice of the geometry.
 
\section{Introduction}
Enlarging spacetime seems to be an active field of research nowadays. Visitors
to the arXiv server from Cornell will be treated to an abundance of wildly
diverging theories which involve extra dimensions varying in number and size.
At present it is not clear which of these models will succeed in the end.
With experimental capabilities being limited a definite answer will certainly
take time. The basic idea, which originated decades ago in work by Kaluza [1]
and Klein [2], is likely to haunt our dreams for many more years with no
prospect of ending any time soon. At the same time the physicist must keep
pace with an ever growing mathematical literature concentrating solely on 
algebraic and geometrical issues. And with it grows the risk of frustration.

While the initial motivation for studying extra dimension was the unification
of gravitation and electromagnetism, many more reasons why extra dimensions
ought to be taken seriously have been stated lateron leading to a revival
of interest and research in this field. At issue, however, is more than just
the ``bottom up approach'' to unification. Notably, string theory
striving for quantum gravity
can consistently be formulated in spaces with six or seven extra
dimensions. These extra dimensions, suitably compactified on the Planck scale,
remain unobservable in laboratory experiments. Though unobservable, the 
proposed structure is carefully motivated and provides new physical insight.
The impact of string theory on our way of thinking cannot be overestimated.

Research has exceedingly been activated by ideas put forward by
Arkani-Hamed, Dimopoulos, and Dvali [3] who addressed the Higgs mass hierachy
problem by introducing \emph{large} extra dimensions at millimeter level.
The proposed models were immediately brought in contact with superstrings
[4], while soon after Randall and Sundrum [5] proposed warped large extra
dimensions, also aiming at solving the hierarchy problem. Here
the internal manifold should become visible if one probes the model at short
distances of the order given by the (macroscopic) compactification scale.
Indeed, it has been suggested to probe such a scenario with neutrinos [6][7].

The idea that spacetime is but a surface embedded in some higher dimensional
space has a long history. See for instance [8] and [9]. It leads to the 
concept of braneworlds. While fermions live on the brane and even gauge
fields are prevented from propagating away from it, gravitons extend into 
the bulk. The major problem here is to obtain 4-dimensional gravity on the
brane. Properties of D-branes in the context of closed strings have been
discussed by Polchinski [10].

In addition, theories have been proposed where the internal space is no
longer compact but has an infinite volume [11] to solve the problem of
the cosmological constant [12] [13] and the accelerating Universe.
For a review see the ICTP Lectures by Gabadadze [14].

\section{Motivation}

As a gauge theory the Standard Model of elementary particle physics is
based on some Lie group $G$ with Lie algebra
\begin{equation}
 \mbox{Lie\,}G \cong{\bf su}(3)\oplus{\bf su}(2)\oplus{\bf u}(1)\ . \label{Lie}
\end{equation}
and in [15] it was explained why the choice
\begin{equation}
    G=S(U(3)\times U(2))  \label{group}
\end{equation}
is more than an educated guess, but seems preferred from the physics point 
of view. The group $G$ is connected but not simply connected: its fundamental
group is $\pi_1(G)=\ZZ$. The embeddings
\begin{equation}
  \label{eq:emb}
   G\ \subset\ SU(5)\ \subset\ U(5)\ \subset\ SO(10)\ \subset\ E_8
\end{equation}
may be relevant to model building if one wants to go beyond the Standard Model.
With $G$ acting on $\CC^5$ one then finds 
that the exterior algebra $\bw \CC^5$, considered as a complex linear
space of dimension $2^5$, ist best suited to unify leptons and quarks
(together with their antiparticles) of one generation in a single, albeit 
highly reducible representation of the gauge group $G$. Again, we refer
to [15] for details and the absence of anomalies.

Though we see faintly and partly the mathematical structure, the physical
role of the space $\CC^5$ remains mysterious. From the huge set of
possible representation spaces, why does nature choose the space $\bw \CC^5$
which is neither irreducible under $G$ nor under the larger group $SU(5)$, 
the candidate for grand unification?
Does $\CC^5$ have its origin in geometry
and an interpretation in terms extra dimensions? The idea we want to pursue
borrows from the theory of spinor bundles. For this purpose let the physical 
space be a direct product
\begin{eqnarray}
  \label{eq:space}
                 M=M_1\times M_2
\end{eqnarray}
of two manifolds where $M_1$ is the 4-dimensional spacetime, 
visible on
large length scales, whereas $M_2$, often called the \emph{internal
manifold}, arises from a compactification of extra dimensions on small
scales. For laboratory ex\-pe\-ri\-ments it suffices to assume that $M_1$
coincides with its tangent space, the Minkowski space. Thus, bundles on $M_1$
will be trivial, and all subtleties and difficulties are encountered in the
internal manifold. Let $M_2$ carry
a Riemannian metric and let its dimension be even, say $2n$. 

The Clifford algebra concept enters naturally into Riemannian geometry
and leads to basic topics like spin geometry und Dirac operators. Focussing
on the cotangent bundle leads to a number of constructions. For instance,
one may associate a Clifford algebra to each point $p\in M_2$ and its 
cotangent space. The collection of these algebras then forms the
Clifford bundle $C(M_2)$. The concept of spinors may be generalized,
i.e., there is, at each point $p\in M_2$, a complex linear space $S_p$ of 
dimension $2^n$ on which the Clifford algebra of that point acts irreducibly.
However, there may be a topological obstruction to construct spinor spaces
globally and smoothly so as to get a 
spinor bundle $\mathfrak{S}_2$. Hence, we need a further
assumption: $M_2$ has a \emph{spin structure}.

We adhere to the conception that fundamental fermions are described by
some spinor field $\psi$ where ``spinor'' refers to the underlying manifold,
in our case the direct product of $M_1$ and $M_2$, and thus can be thought 
of as an element of the tensor product
\begin{equation}
  \label{eq:spin}
         S= S_1\otimes S_p\ ,\qquad \mbox{dim\,}S=2^2\cdot 
        2^n=2^{\mbox{\footnotesize dim\,}M/2}
\end{equation}
where $S_1$ is the complex linear space of Dirac spinors obtained from the Minkowski metric
on $M_1$. To accommodate leptons and quarks of one generation in a single
spinor, we need only identify $S_p$ with $\bw \CC^5$, thus assuming 
$n=5$. This forces us to require that the internal manifold $M_2$ be 
10-dimensional. The present setting leads to a change of our views on
fundamental fermions and suggests a
new and unusual interpretation of their degrees of freedom.
For instance, what distinguishes leptons from quarks, what we call color
and weak isospin is now thought of as some sort of \emph{polarisation} 
referring
to the internal manifold, quite similar to the notion of spin polarisation 
which relates to Minkowski spacetime.

The isomorphism $S_p\cong\bw \CC^5$ should not come as a surprise. 
It is nobody's
secret that every spinor space arises as the underlying linear space of
some complex exterior algebra. Therefore, its dimension is always a power
of two. Such point of view leads us to introduce the real-linear 
10-dimensional Euclidean space $E_{10}$ with some complex structure, i.e., 
there is some real orthogonal transformation $J:E_{10}\to E_{10}$ satisfying 
$J^2=-\mbox{id}$. Consequently, $J$ has eigenvalues $\pm i$ in $E_{10}\otimes\CC$ and 
eigenspaces $V$ and $V^*$ of equal dimension such that
$$ E_{10}\otimes\CC=V^*\oplus V,\qquad J=i\ (-i)\ \ \mbox{on}\ V\ (V^*)\ . $$
Given some basis in $V$ we may identify $V$ with $\CC^5$ and $\bw V$ with
$\bw\CC^5$. The above decomposition of the complexified 
$E_{10}$ is called a \emph{polarisation} of $E_{10}$.
It is characteristic of a polarisation that the spaces $V$ and $V^*$ 
are isotropic
in the following sense. The Euclidean metric on 
$E_{10}$ extends to some complex bilinear form $(\ ,\ )$ on $E_{10}\otimes\CC$
such that $(v,v)=0$ whenever $v\in V$ or $v\in V^*$. 

As suspected, polarisations of $E_{10}$ are abundant. There are as many 
polarisations as there are complex structures on $E_{10}$. If we have made
one such choice for $J$, then conjugating $J$ by elements of $SO(10)$ will
produce all other choices. Also, conjugation by elements of $U(5)$ leaves
the previous choice unaltered. Hence, the set of all all complex structures
is naturally parametrized by the coset space $SO(10)/U(5)$. Since our
objective is to obtain a \emph{canonical polarisation}, we assume $J$ to
be fixed once and for all.

One takes $S_2=\bw V$ as the spinor space associated to the 
Euclidean space $E_{10}$ and justifies this definition
by giving $S_2$ the structure of an irreducible
Clifford module, in fact a standard two-steps process. First, one constructs
the real Clifford algebra $C(E_{10})$ over $E_{10}$ with canonical injection
$E_{10}\to C(E_{10})$. Second, one provides a
Clifford map\footnote{This is a map satisfying 
             $c(v)^2=-(v,v)$ for all $v\in E_{10}$\ .}
$$                  E_{10}\quad\arr{c}\quad \mbox{End\,}S_2\ . $$
As it turns out, the canonical extension
$$               C(E_{10})\otimes\CC\quad\arr{c}\quad\mbox{End\,}S_2 $$ 
is an algebraic isomorphism. Details of this construction may be found
in the mathematical literature [25]. For a short account, mainly to serve 
the needs of physicists, see [26].

Once we accept the idea of having as many as ten extra 
dimensions, the afore mentioned mystery has been 
unveiled. The message seems to be that the space $\bw\CC^5$,
used to describe the internal degrees of freedom of each fermion generation,
is in fact a spinor space, and $\CC^5$ should be thought of as the positive 
part of a polarisation of $T^*_pM_2$, the cotangent space of the internal 
manifold at a given point $p\in M_2$. The polarisation is canonical whenever 
$T^*_pM_2$ carries a complex structure. If there is a bundle automorphism
$J:TM_2\to TM_2$ satisfying $J^2=-\mbox{id}$, the manifold is said to carry an 
\emph{almost complex structure}. Since each cotangent space inherits
the complex structure, our previous analysis applies.

It is tempting to assume that the internal manifold $M_2$ is
a K\"ahler manifold. We remind the reader that a Riemannian metric
$(\ ,\ )$ on $M_2$ is said to be K\"ahlerian if $J$ is pointwise orthogonal,
that is we have
$$         (v,w)_p= (Jv,Jw)_p\ ,\qquad v,w\in T_pM_2 $$
at all points $p\in M_2$, and if $\nabla J=0$ where $\nabla$ denotes the
Levy-Civita connection of the underlying Riemannian structure. It is known
that the almost complex structure of a K\"ahler manifold is always integrable,
meaning it comes from a system of local complex coordinates which are
holomorphically related (i.e., transition functions between
different coordinate patches are holomorphic). Among the variety of 
compact complex 
manifolds carrying a spin structure we find the projective spaces $\CC P_n$
provided the complex dimension $n$ is odd. The projective 3-space $\CC P_3$
has been used in Penrose's twistor approach to the analysis of field
equations on spacetime. The projective 5-space $\CC P_5$ might be a
candidate for our purposes.
It is also known that any compact complex
manifold, which can be embedded in $\CC P_n$ for some $n$, is in fact a
K\"ahler manifold.

Embedded in the Clifford algebra $C(E_{10})$ is a simply
connected Lie group
$$   \mbox{Spin}(10) =\mbox{Spin}\,E_{10}$$
called the \emph{spin group} associated with the Euclidean space.
The canonical injection $E_{10}\to C(E_{10})$ allows us to think of
the Euclidean space as part of the algebra.
The spin group acts on $C(E_{10})$, hence on $E_{10}$ through the adjoint 
representation.
As the action preserves the Euclidean structure, we have a short exact sequence
$$ 1\ \to\ \ZZ_2\ \to\ \mbox{Spin}(10)\ \to\ SO(10)\ \to\ 1$$
($\ZZ_2=\{\pm 1\}$) telling us that $\mbox{Spin}(10)$ 
is the universal covering group of the orthogonal group $SO(10)$. 
Though the Clifford algebra acts irreducibly
on the spinor space $S_2$, the spin group does not. It will later be shown how
$\bw \CC^5$, the unifying space of quarks and leptons, decomposes under the 
action of $\mbox{Spin}(10)$. It may then be interesting to learn that
the spin group mixes leptons and quarks, particles and antiparticles, but
respects the chirality. We also remind the reader that, with regard to the
external space, the spin group is $SL(2,\CC)$ acting reducibly on $S_1$,
the space of Dirac spinors. If, however, one prefers to work with the
Euclidean 4-space in place of the Minkowski space, as is often the case in
field theory, then the spin group would be the compact group 
$SU(2)\times SU(2)$.

Our assumption that there be a spin structure means that there exists
a principle $\mbox{Spin}(10)$-bundle $\mbox{Spin}(M_2)$ such that
the cotangent bundle arises as an associated bundle with
$E_{10}$ as fiber space:
$$  T^*M_2=\mbox{Spin}(M_2)\times_{\mbox{\scriptsize{Spin}}(10)}E_{10}\ .$$
Among the consequences is the observation that the manifold $M_2$ is oriented
with frame bundle
$$   SO(M_2)=\mbox{Spin}(M_2)\times_{\mbox{\scriptsize{Spin}}(10)}SO(10)$$
which doubly covers the principle $\mbox{Spin}(10)$-bundle. 
Many other bundles are
constructed as associated bundles. For instance, the Clifford bundle may now
be written as
$$   C(M_2)=\mbox{Spin}(M_2)\times_{\mbox{\scriptsize{Spin}}(10)}C(E_{10})\ .$$
With the spinor space $S_2$ carrying an irreducible representation of 
$C(E_{10})$, there is an implied irreducible action of the Clifford bundle 
on the spinor bundle
\begin{equation}
  \label{eq:spinb}
  \mathfrak{S}_2 =\mbox{Spin}(M_2)\times_{\mbox{\scriptsize{Spin}}(10)}S_2\ .
\end{equation}
Unfortunately, it may happen that there exist several inequivalent spin 
structures on $M_2$, hence different principal bundles $\mbox{Spin}(M_2)$.
However, if $M_2$ is simply connected (as will be assumed), the spin structure
is unique.

\section{$\ZZ_2$-Grading and Chirality}

To this point we have paid little attention to the structure of spinor
spaces although we expect the reader to be familiar with the notion of
\emph{chirality}, i.e., the physical interpretation of the eigenvalues
$\pm 1$ of the $\gamma_5$ matrix acting on $S_1$, the complex four-dimensional
space of Dirac spinors. Abstractly speaking we may say that $S_1$ is
$\ZZ_2$-graded:
\begin{equation}
  \label{eq:grad1}
  S_1=S_1^+\oplus S_1^-\ ,\qquad \gamma_5=\pm 1 \ \mbox{on}\ S_1^\pm\ .
\end{equation}
If a spinor belongs to $S_1^\pm$, it has a definite chirality and is called 
a \emph{Weyl spinor}. Weyl 
spinors in $S_1^+$ are said to be right-handed, those in $S_1^-$ left-handed.
One of the most striking phenomena in particle physics is the fact that
the fundamental fermion fields entering the Lagrangian are grouped according
to their chirality, that is they take values either in $S_1^+$ or in $S_1^-$.

Our aim is to demonstrate that there is another chirality operator on $S_2$,
the analogue of $\gamma_5$, giving $S_2$ a $\ZZ_2$-graded structure,
and to explore the physical relevance of this fact. Namely,
with respect to an orthonormal basis $\{e_i\}_{i=1}^{10}$ in $E_{10}$ 
we define the chirality operator as
$$   \Gamma =ie_1e_2\cdots e_{10}\ \in C(E_{10})\otimes\CC\ .$$
The factor $i$ in front has been chosen so as to guarantee that $\Gamma^2=1$.
It is only for this factor that we passed to the complexified Clifford
algebra.
The question arises whether the definition of $\Gamma$ depends on our choice
of the basis. Suppose  $\{e'_i\}_{i=1}^{10}$ is another orthonormal basis in
$E_{10}$ and
$$   \Gamma' =ie'_1e'_2\cdots e'_{10}\ .$$
Then there exists a matrix $A\in O(10)$ that relates the new basis to the 
first one, and it is quickly realized that $\Gamma'=(\mbox{det}A)\Gamma$
where $\mbox{det}A=\pm 1$. Given some orientation of $E_{10}$ and requiring
that all bases be oriented, there will be no ambiguity since $A$ is then 
restricted to $SO(10)$.

Recall now that $E_{10}$ is assumed to carry a complex structure.
We may thus choose a basis satisfying 
$$     Je_k=\cases{\phantom{-}e_{k-1}& if $k$ is even\cr
                             -e_{k+1}& if $k$ is odd}    $$ 
which gives the Euclidean space a canonical orientation. Moreover,
\begin{equation}
  \label{eq:basV}
      \e_k= \frac{1}{\sqrt2}(e_{2k-1}+ie_{2k})\qquad(k=1,\ldots,5)
\end{equation}
becomes a basis in $V$. The induced basis $\e_I$ in $\bw V$ is given by
$$ \e_I=\e_{i_1}\wedge\cdots\wedge\e_{i_k}\in\bw^kV,
\qquad I=\{i_1,\ldots,i_k\},\quad i_1<\cdots<i_k,\quad 0\le k\le 5$$
where $I$ runs over all subsets of \{1,2,3,4,5\} including the empty set.
This establishes the isomorphisms $V\cong\CC^5$ and $\bw V\cong\bw\CC^5$.

The action of $\Gamma$ on the spinor space $S_2$ leads to a $\ZZ_2$-graded
structure,
\begin{equation}
  \label{eq:s2gr}
  S_2=S_2^+\oplus S_2^-\ ,\qquad \Gamma=\pm 1\ \mbox{on}\ S_2^\pm\ ,
\end{equation}
quite analogous to the decomposition $S_1=S_1^+\oplus S_1^-$ for the space of
Dirac spinors. Before showing its physical significance we point at an
apparent dilemma. For there seem to be two conflicting $\ZZ_2$-structures on 
$S_2$, one given by the chirality operator $\Gamma$ and another one induced 
by the exterior algebra structure:
$$
  S_2=\bw^+V\oplus\bw^-V\ ,\qquad 
      \bw^+ V=\sum_{k=\mbox{\scriptsize even}}\bw^kV\ ,\qquad 
      \bw^- V=\sum_{k=\mbox{\scriptsize  odd}}\bw^kV
$$
It may be checked, however, that these two gradings coincide, i.e.,
$S_2^\pm=\bw^\pm V$. This in particular implies that the spaces $S_2^\pm$
have same dimension and so
$$  \mbox{dim}\,S_2^\pm =2^5/2 =16 \ .$$
For a proof see [26].

We continue by pointing at the $\ZZ_2$-graded algebra structure of 
$C(E_{10})$, that is we have the decomposition
$$   C(E_{10})=C^+(E_{10})\oplus C^-(E_{10}) $$
into subspaces of equal dimension and
$$  C^+(E_{10})\,C^\pm(E_{10})\subset C^\pm(E_{10})\ ,\qquad
    C^-(E_{10})\,C^\pm(E_{10})\subset C^\mp(E_{10})\ .
$$
To be more specific, $C^\pm(E_{10})$ is created by even resp.\ odd powers
of its generators $e_k$, and $C^+(E_{10})$ includes the real numbers
as a subalgebra. We also point at the obvious fact that $C^+(E_{10})$ is a
subalgebra while $ C^-(E_{10})$ is not, and 
$$          \mbox{Spin}(10)\subset C^+(E_{10})\ .$$
Furthermore, the action of the Clifford algebra on the spinor space respects
the grading in the sense that
$$      C^+(E_{10})S_2^\pm\subset S_2^\pm\ ,\qquad
        C^-(E_{10})S_2^\pm\subset S_2^\mp\ .   $$
This follows easily from the observation that the chirality operator
$\Gamma$ anticommutes with the generators $e_k$.
Since $S_2^+$ and $S_2^-$ are invariant subspaces under $C^+(E_{10})$,
they are also invariant under $\mbox{Spin}(10)$. It is a general fact that,
for models based on an even dimension (10 in our case), the
representation of the spin group on the spinor space commutes with the
chirality operator.

To summarize, there is a 16-dimensional representation of $\mbox{Spin}(10)$
on $S_2^+$ and another one on $S_2^-$. Both representations are irreducible.
Since the action of $\mbox{Spin}(10)$ on $S_2$ respects its grading, we may
define
$$   \mathfrak{S}^\pm_2=\mbox{Spin}(M_2)\times_{\mbox{\scriptsize Spin}(10)}
     S_2^\pm $$
which gives the spin bundle the induced graded structure.

\section{Avoiding Mirror Fermions}

There is an obvious and simple prescription for constructing tensor products 
of graded
spaces giving the product an induced grading. Applying this prescription
to the spinor space $S=S_1\otimes S_2$, connected with the manifold
$M=M_1\times M_2$, leads to the grading $S=S^+\oplus S^-$, where
\begin{eqnarray*}
    S^+ &=& (S_1^+\otimes S_2^+)\oplus(S_1^-\otimes S_2^-)\\
    S^- &=& (S_1^+\otimes S_2^-)\oplus(S_1^-\otimes S_2^+)\ ,
\end{eqnarray*}
and to yet another concept of ``chirality'' or ``parity'', inferred from
the $\ZZ_2$-graded structure of the total spinor space $S$.
How will this new type of parity of elementary fermions manifest itself?
By appeal to experience (see [15] for details), fermions in $S_2^+$
are right-handed while those in $S_2^-$ are left-handed. So their Dirac
spinors lie in $S_1^+$ and $S_1^-$ respectively, meaning that fundamental
fermions are confined to $S^+$. We emphasize that this follows solely from 
the existing experimental data.

If we were to introduce yet another family of fermions with degrees of freedom
described by the complement $S^-$, it would mean that we allow for so-called
\emph{mirror fermions}. The occurrance of mirror fermions is frequent feature
of family unification schemes, especially when $E_8$ is taken as the 
unification group [16]. It is known, however, that the presence of mirror
families is in conflict with experiment and also leads theoretical
difficulties [17].

Placing all fundamental fermions into the subspace $S^+\subset S$ does not
imply that its complement $S^-$ is irrelevant, i.e., plays no role in the
Standard Model as a field theory. For there are operators of odd type in the
theory, like the Dirac operator, that pass from $S^+$ to $S^-$ and vice
versa.

Let us now look at the two subspaces of $S^+$ and the fermions they describe,
restricting attention to the first generation. We place 16 right-handed
Weyl spinors into $S_1^+\otimes S_2^+$ corresponding to 8 matter fields
$$           (\nu_e,e,u,d)_R $$
and 8 antimatter fields\footnote{The superscript ${}^c$ means passage to the
conjugate field (changing from particles to antiparticles).}
$$           (\nu_e^c,e^c,u^c,d^c)_R $$
(counting the color degrees of freedom of $u$ and $d$ quarks). We then place 16 left-handed
Weyl spinors into $S_1^-\otimes S_2^-$ which also correspond to 8 matter fields
$$           (\nu_e,e,u,d)_L $$
and 8 antimatter fields
$$           (\nu_e^c,e^c,u^c,d^c)_L \ .$$
Effectively, we have placed all 16 right-handed (anti)matter fields into
one irreducible representation of the product spin group
$$             SL(2,\CC)\times\mbox{Spin}(10) $$
and all 16 left-handed (anti)matter fields into another irreducible
representation of the same group. This is almost in accordance with previous
$SO(10)$ grand unification models [18] [19], except that we have chosen
a slightly different assignment of particles and Weyl fields leaving
their quantum numbers unchanged\footnote{Quark-lepton unification has a long
history. The first attempt in this direction was made by Pati and Salam [20]
and independently by Georgi and Glashow [21] in 1974. In the latter proposal
the fermions in $\{5\}=\bw^1\CC^5$ were chosen right-handed, those in
$\{10\}=\bw^2\CC^5$ left-handed. The trivial representation of $SU(5)$ on 
$\{\bar{1}\}=\bw^5\CC^5$  corresponding to the right-handed neutrino entered 
the scene one year later [22][23] when $SO(10)$ was the group of choice.
From then on it was understood that all 16 left-handed fundamental Weyl 
spinors of the Standard Model form the space $\{1\}+\{10\}+\{\bar{5}\}=
\bw^0\CC^5\oplus\bw^2\CC^5\oplus\bw^4\CC^5=\bw^+\CC^5$, whereas we decided
to put all 16 \emph{right-handed} Weyl fields in that space.}.

For the convenience of the reader we shall now recall the formalism and 
some of the results
of [15] where the assumption was that the symmetry group is 
$G=S(U(3)\times U(2))$. As a subgroup of $SU(5)$ it acts naturally on 
$$         \bw\CC^5=\bw(\CC^3\oplus\CC^2)=\bw\CC^3\,\otimes\,\bw\CC^2$$
with irreducible subspaces $\bw^p\CC^3\otimes\bw^q\CC^2$ labelled by
$$    p\in\{0,1,2,3\} \qquad q\in\{0,1,2\}\ .$$
Their dimensions are given as $({3\atop p})({2\atop q})$. Leptons, identified
as color singlets, have $p=0$ or 3, while quarks, being color triplets, have
$p=1$ or 2. On the other hand, the group $SU(2)$ of weak isospin generates
singlets ($q=0$ or 2) and doublets ($q=1$). One of the outstanding features 
of the Standard Model is that $p$ and $q$ already determine the chirality 
as $(-1)^{p+q}$, for we have
$$  \bw^+\CC^5=\sum_{p+q=\mbox{\scriptsize even}}
    \bw^p\CC^3\otimes\bw^q\CC^2\ ,\qquad
    \bw^-\CC^5=\sum_{p+q=\mbox{\scriptsize odd}}
    \bw^p\CC^3\otimes\bw^q\CC^2\ .$$
The hypercharge obeys the fundamental relation
$$           Y=\frac{2}{3}p-q $$
giving fractional values to quarks and integer values to leptons. Note that,
in this discussion, we have treated matter and antimatter on the same
footing. Nevertheless, there is a parity which distinguishes these two:
$$   p=\mbox{even:\ \ matter}\ ,\qquad p=\mbox{odd:\ \ antimatter}\ .$$ 
Note also that the gauge group $G$ acts trivially on the one-dimensional spaces
$\bw^0\CC^5$ and $\bw^5\CC^5$ connected to the right-handed neutrino
and left-handed antineutrino respectively. Consequently, the Weyl fields
involved do not interact with the gauge fields.  

\section{The Fundamental Fermion Field Operator}

Needless to say that our interest lies in the Standard Model as a
quantum field theory rather than in the pre-quantized version of it. The 
problem we therefore face is this. While the assumed first-generation fermion 
field $\psi_1(x)$ is a \emph{fixed} operator, all that can vary is the state
$\Phi$, member of a one-particle Hilbert space $H$. So far we have been rather
vague about the connection of the operator, the state, and the spinor space.
The following should be clear. What assumes different values in $S^+$ is not 
the field $\psi_1(x)$ but rather the wave function $\Phi(x)$ (a spinor) of the
one-particle state obtained from $\psi_1$ as a ``matrix element'':
$$      \Phi(x)= (\Omega,\psi_1(x)\Phi)\ \in\ S^+ $$
where $\Omega$ represents the vacuum. As we vary $x\in M_1$ and $\Phi\in H$,
any spinor $\Phi(x)\in S^+$ can be reached. In a pre-quantized formulation
of particle physics one deals with the one-particle spinor wave functions 
$\Phi(x)$ exclusively. 

It is common practice to assume global gauge transformations
$g\in G$ to be unitarily implemented in the Fockspace $\overline{\bw H}$
(the completion of the exterior algebra over $H$) such that, 
for all $g\in G$ and $\Phi,\Phi'\in \overline{\bw H}$,
$$    (U(g)\Phi',\psi_1(x)U(g)\Phi)=D(g)(\Phi',\psi_1(x)\Phi)$$
where $g\mapsto D(g)$ is the relevant matrix representation of $G$ on $S^+$.
The latter property is equivalently stated as a covariance property of the
field operator,
$$   U(g)^{-1}\psi_1(x)U(g)=D(g)\psi_1(x)\ ,$$
which ultimately justifies the usage \emph{the field $\psi_1$ takes values in}
$S^+$, and by abuse of notation we shall write $\psi_1(x)\in S^+$.
More generally, we adopt the following usage.
\begin{quote}
  \em If a field is said to be section of a 
      vector bundle, this will be short for saying that
      all matrix elements with respect a dense set of states are
      smooth sections.
\end{quote}
The most striking feature of Standard Model is the triple repetition 
of a concept called \emph{generation}. Various models beyond the Standard
Model have been proposed
to explain why there are precisely three generations, also called families
[24]. Here we shall not pursue the idea that the family structure arises
from a large group representation but assume that it comes from topology.

In a conventional formulation one would work with $3\cdot16$ fermion
fields, formally sections of the trivial spinor bundle 
$$        \mathfrak{S}_1=M_1\times S_1\ .$$
As Weyl fields they are in fact sections of either $\mathfrak{S}_1^+$ or
$\mathfrak{S}_1^-$.
Then one groups family members (plus their charge conjugates) together 
to obtain three fields taking values in $S^+$, each field
describing one family. The next logical step towards family unification is
to introduce but one field $\psi$ referred to as the
\emph{fundamental field}, formally a section of the positive part of
the combined spinor bundle over $M$,
$$               \mathfrak{S}^+=(\mathfrak{S}_1\otimes\mathfrak{S}_2)^+\ .$$
In less abstract terms it means
$$    \psi(x,p)\in (S_x\otimes S_p)^+\qquad (x\in M_1,\ p\in M_2)$$
where $S_x=S_1$ for all $x$ as the bundle $\mathfrak{S}_1$ is trivial,
and $S_p$ is the fiber of the spinor bundle $\mathfrak{S}_2$ at the point 
$p\in M_2$. Note that, though $S_p$ and $S_2$ are isomorphic, there exists
no canonical isomorphism between the two if the spinor bundle 
$\mathfrak{S}_2$ is
non-trivial. However, given a section of the frame bundle $SO(M_2)$
(consistent with the complex structure), it will
at once lead to the identification
$$            S_p=S_2=\bw\CC^5\ .$$
No attempt is made here to write down an action integral for $\psi$, nor do we
offer an explanation how the known particles arise as exitations with a
huge hierarchy of masses. We have
a vague idea though for what might be the origin of the family structure:
each family corresponds to some fixed point $p\in M_2$ of some automorphism 
group of the internal manifold. 

\section{K\"ahler, Calabi-Yau and Beyond}

The best understood complex manifold is $\CC^n$ though it is embarrassingly 
simple.
At each point, $dz_1,\ldots,dz_n$ form a basis for the cotangent space.
With its standard metric
  \begin{equation}
    \label{eq:s1}
     g =|dz_1|^2+\cdots+|dz_n|^2\ ,
  \end{equation}
the related K\"ahler $(1,1)$-form
\begin{equation}
    \label{eq:s2}
    \omega=\frac{i}{2}(dz_1\wedge d\bar{z}_1+\cdots+dz_n\wedge d\bar{z}_n)
\end{equation}
and the holomorphic volume $(n,0)$-form
\begin{equation}
    \label{eq:s3}
    \Omega=dz_1\wedge\cdots\wedge dz_n
\end{equation}
the space $\CC^n$ provides the simplest example of a K\"ahler manifold
with holo\-no\-my group $SU(n)$. Generally speaking, the holonomy group
is a global invariant of a Riemannian $2n$-manifold. It is a subgroup of
$O(2n)$, or of $SO(2n)$ if the manifold is oriented. Generic K\"ahler
metrics on a complex manifold have holonomy $U(n)\subset SO(2n)$, 
and a Calabi-Yau $n$-fold may be characterized as a compact K\"ahler 
manifold with holonomy $SU(n)$. See [27] for details.

The question arises whether there is a geometry beyond K\"ahler and 
Calabi-Yau, whose holonomy group is the gauge group (\ref{group}) of 
the Standard Model. For this purpose let us write
$$    dz_I =dz_{j_1}\wedge\cdots\wedge dz_{j_r}\ ,\qquad |I|=r$$
whenever $I=\{j_1,\ldots,j_r\}$ and $1\le j_1<\cdots<j_r\le n$. In
addition to (\ref{eq:s1}), (\ref{eq:s2}), and (\ref{eq:s3}) we require that
the $(n-1,n-1)$-form
\begin{equation}
    \label{eq:s4}
    \Theta =\sum_{|I|=n-1} \sigma_I \,dz_I\wedge d\bar{z}_I 
\end{equation}
be invariant, where $\sigma_I$ is the sign of the permutation taking
$\{I^c,I\}$ to its normal order $\{1,2,\ldots,n\}$ and $I^c$ is the
complement of $I$. As $|I^c|=1$ in (\ref{eq:s4}), $I^c=\{j\}$ for some
$j\in\{1,\ldots,n\}$ and $\sigma_I=(-1)^j$. Let
$$   dz_k'=u\,dz_k\ ,\qquad u \in GL(n,\CC)\qquad(k=1,\ldots n) $$ 
be a change of the basis and consider the induced change of $\Theta$:
$$    \Theta' =\sum_{|I|=n-1} \sigma_I dz_I'\wedge d\bar{z}_I'\ .  $$      
From the defining relation
$$ u^cdz_{k_1}\wedge dz_{k_2}\wedge\ldots\wedge dz_{k_n}=
   dz_{k_1}\wedge udz_{k_2}\wedge\ldots\wedge udz_{k_n}$$
for the \emph{cotranspose}\footnote{This familiar concept in algebra comes 
under various names. The name \emph{cotranspose} is taken from Bourbaki [28].}
 $u^c=(u^c_{ji})$ and
$u^cdz_i=\sum_j u^c_{ji}dz_j$ we obtain
$$  dz_i\wedge\Theta'\wedge d\bar{z}_k=
    u^cdz_i\wedge\Theta\wedge \bar{u}^cd\bar{z}_k=
    \left(\textstyle\sum_ju_{ji}^c(-1)^{j+n-1}\bar{u}^c_{jk}\right)\Omega\wedge\bar{\Omega}\ .$$
If $u\in GL(n,\CC)$ preserves $\Theta$, then
$$   
  \sum_j\bar{u}^c_{jk}(-1)^j u_{ji}^c=(-1)^k\delta_{ik}$$
or, written in a matrix notation,
\begin{equation}
  \label{Q}
                  u^{c*}Qu^c =Q\ ,\qquad Q_{ik}=(-1)^k\delta_{ik}\ .
\end{equation}
The Theorem of Laplace, $u^cu= uu^c=(\mbox{det}\,u)\one$, allows to rewrite 
the above condition as
\begin{equation}
  \label{eq:q}
  |\mbox{det}\,u|^2\,Q=u^*Qu\ .
\end{equation}
Recall now that $u$ preserves (\ref{eq:s1}), (\ref{eq:s2}), and (\ref{eq:s3})
iff $u\in SU(n)$. As follows from (\ref{eq:q}), it also preserves 
(\ref{eq:s4}) if $u$ commutes
with $Q$ and hence with the projections $P_\pm=\one\pm Q$. 
Together, these conditions describe precisely the group elements
$$ u\in G=\cases{S(U(m)\times U(m))  & if $n=2m$\cr
                 S(U(m+1)\times U(m))& if $n=2m+1$.}   $$
Basis elements with even resp.\ odd indices span invariant subspaces
corresponding to the projections $P_\pm$; e.g.\ for $n=2m+1$, we let
\begin{quote}
  $U(m+1)$ act on $(dz_1,dz_3,\ldots)$ \\
  $U(m)\phantom{+11}$ act on $(dz_2,dz_4,\ldots)$.
\end{quote}
The ideas presented here carry over to Calabi-Yau n-folds with a
$G$-structure. 
\newpage\noindent
In more detail:
\begin{quote}
  \emph{ A $G$-structure on a complex n-fold is a principal $G$-subbundle
         of the frame bundle. Each $G$-structure gives rise to a quadrupel
         $(g,\omega,\Omega,\Theta)$, such that every cotangent space admits
         an isomorphism with $\RR^{2n}$ identifying $(g,\omega,\Omega,\Theta)$ 
         with the standard quadrupel given by (\ref{eq:s1}), (\ref{eq:s2}), 
         (\ref{eq:s3}), and (\ref{eq:s4}).}
\end{quote}
Note that $d\omega=d\Theta=d\Omega=0$ always and
$$   \omega^n=n!(-1)^{n(n-1)/2}(i/2)^n\,\Omega\wedge\bar{\Omega}\ .$$
Finally, it is easily checked that
$$ \omega\wedge\Theta=\cases{-\Omega\wedge\bar{\Omega}& if $n=$odd\cr
                             0 & if $n=$even}   $$
The emerging picture is this: a suitable complex 5-fold within this category 
might play the role of the internal manifold $M_2$.

\vspace{2cm}
\noindent
{\bf References}
\begin{enumerate}
\item Th.\ Kaluza, Sitzungsberichte der Preussischen Akademie der 
Wissen\-schaf\-ten, Phys.Math.Klasse
996 (1921). Reprinted and translated in T.\ Appelquist, A.\ Chodos, and
P.G.O.\ Freund (ed.): \emph{Modern Kaluza-Klein Theories}, Addison-Wesley,
1987.
\item O.\ Klein, Zeits.f.Physik {\bf 37}, 895 (1926). Reprinted and translated
in T.\ Appelquist, A.\ Chodos, and P.G.O.\ Freund (ed.): \emph{Modern 
Kaluza-Klein Theories}, Addison-Wesley, 1987.
\item N.\ Arkani-Hamed, S.\ Dimopoulos, and G.R.\ Dvali, hep-ph/9803315 and
Phys.Lett.B {\bf 429}, 263 (1998). Also hep-ph/9807344 and Phys.Rev.D {\bf
59}, 086004 (1999).
\item I.\ Antiniadis, N.\ Arkani-Hamed, S.\ Dimopoulos, and G.R.\ Dvali, 
hep-ph/9804398 and Phys.Lett.B {\bf 436}, 257 (1998).
\item L.\ Randall and R.\ Sundrum, hep-th/9906064 and Phys.Rev.Lett.\ {\bf
83}, 4690 (1999). Also hep-ph/9905221 and Phys.Rev.Lett.\ {\bf 83}, 3370
(1999).
\item N.\ Arkani-Hamed, S.\ Dimopoulos, G.R.\ Dvali, and J.March-Russell,
hep-ph/9811448 and Phys.Rev.D {\bf 65}, 024032 (2002).
\item  G.R.\ Dvali and A.Y.\ Smirnov, hep-ph/9904211 and Nucl.Phys.B {\bf
      563}, 63 (1999)
\item V.A.\ Rubakov and M.E.\ Shaposhnikov, Phys.Lett.B {\bf 125}, 136 (1983).
\item K.\ Akama, hep-th/0001113 and Lect.Notes Phys.\ {\bf 176}, 267 (1982)
\item J.\ Polchinski, hep-th/9510017 and Phys.Rev.Lett.\ {\bf 75}, 4724
      (1995). Also hep-th/9611050.
\item G.R.\ Dvali, G.\ Gabadadze, and M.\ Porrati, hep-th/0005016 and
      Phys. Lett.B {\bf 485}, 208 (2000).
\item G.R.\ Dvali and G.\ Gabadadze, hep-th/0008054 and Phys.Rev.D {\bf 63},
      065007 (2001)
\item G.R.\ Dvali, G.\ Gabadadze, and M.\ Shifman, hep-th/0202174 and
      Phys. Rev.D {\bf 67}, 044020 (2003). Also hep-th/0208096.
\item G.\ Gabadadze, hep-ph/0308112 and CERN-TH/2003-157.
\item G.\ Roepstorff, hep-ph/0006065, hep-th/0005079 and
      phys.stat.sol.(b) {\bf 237}, 90 (2003).
\item There are many papers on this issue. For a review and a list of
      references see S.L.\ Adler, hep-ph/0201009. 
\item G.\ Triantaphyllou, Int.J.Mod.Phys.A {\bf 15}, 265 (2000).
      J.Phys.G: Nucl. Part.Phys.\ {\bf 26}, 99 (2000). hep-ph/9908251.
      hep-ph/0109023. Mod. Phys.Lett.A {\bf 16}, 53 (2001).  
\item A.\ Zee, \emph{ Unity of Forces in the Universe}, Vol.1, Sec.IV,
      World Scientific 1982.
\item R.N.\ Mohapatra, \emph{Unification and Supersymmetry}, Chapter 7,
      3rd Ed., Springer 2003
\item J.C.\ Pati and A.\ Salam, Phys.Rev.D {\bf 10}, 275 (1974)
\item H.\ Geogi and S.L.\ Glashow, Phys.Rev.Lett.\ {\bf 32}, 438 (1974)
\item H.\ Georgi, in \emph{Particles and Fields}, edited by C.E.\ Carlson,
      Am.\ Inst.\ Phys.\ 1975
\item H.\ Fritzsch and P.\ Minkowski, Am.Phys.\ {\bf 93}, 193 (1975)
\item P.\ Ramond, hep-ph/9809459. See also reference [19].
\item C.\ Chevalley, \emph{The Algebraic Theory of Spinors}, Columbia
      University Press, New York 1954.\\
      H.B.\ Lawson and M.-L.\ Michelson, \emph{Spin Geometry},
      Princeton University Press, Princeton N.J. 1989.
\item G.\ Roepstorff and Ch.\ Vehns, math-ph/9908029.
\item M.\ Gross, D.\ Huybrechts, and D.Joyce, \emph{Calabi-Yau Manifolds
      and Related Geometries}, Lecture Notes, Springer 2003   
\item N.\ Bourbaki, \emph{Elements of Mathematics, Algebra I}, Chapter III,
      \S8.6, Hermann, Paris 1974  
\end{enumerate}
\end{document}